\def\be{\begin{equation}}
\def\bea{\begin{eqnarray}}
\def\eea{\end{eqnarray}}
\def\ee{\end{equation}}
\def\tj{\theta_{\rm j}}

\def\gmin{\Gamma_{\rm min}}

\def\Rph{R_{\rm ph}}

\def\tj{\theta_{\rm j}}

\def\g0{\Gamma_{0}}

\documentclass{ws-ijmpd}
\usepackage[super,compress]{cite}
\usepackage{hyperref}
\begin{document}

\markboth{Vyas and Pe'er}
{Photon scattering in shearing plasma}

%
\catchline{}{}{}{}{}
%

\title{Theory of photon scattering in shearing plasma: Applications to GRBs 
}

\author{Mukesh Kumar Vyas
}

\author{Asaf Pe'er}

\address{Bar Ilan University, \\
Ramat Gan, \\
Israel, 5290002  \\
mukeshkvys@gmail.com, asaf.peer@biu.ac.il}


\maketitle

\begin{history}
\received{}
\revised{}
\end{history}

\begin{abstract}
We explore a new mechanism for photon energy gain in a relativistic plasma with velocity shear. This process takes place in optically thick plasma and resembles conventional Fermi acceleration, where photons undergo multiple scatterings between regions with varying Lorentz factors, leading to an overall energy increase. The resulting high-energy spectra from the escaped photons exhibit a power-law form. The mechanism is an alternative to the classical radiation spectrum from power-law accelerated particles, which can produce power-law spectra in sources like Gamma-ray bursts (GRBs) and Active Galactic Nuclei (AGNs). By employing both numerical simulations and theoretical analysis, we calculate the expected spectra for GRBs, and show that they match the observed photon indices ($\beta$) at high energies.
\end{abstract}


\ccode{PACS numbers:}


\section{Introduction}
Power-law spectra are commonly observed in astrophysical sources such as gamma-ray bursts (GRBs) \cite{{Band.etal.1993,Kaneko.etal.2006ApJS..166..298K,Bosmjak.etal.2014A&A...561A..25B,pe'er2015AdAst2015E..22P,Barraud.etal.2003A&A...400.1021B}} and active galactic nuclei (AGNs) \cite{Nandra&Pounds1994, Page.etal.2005}. These spectra are generally attributed to synchrotron radiation \cite{{Band.etal.1993,Kaneko.etal.2006ApJS..166..298K}}, where accelerated charged particles emit in the presence of magnetic fields, or to thermal Comptonization \cite{1970RvMP...42..237B,1986rpa..book.....R,1993ApJ...409L..33Z, vyas.etal.2021.predictingAPJL}. In these cases, particles are thought to be accelerated into a power-law distribution through a Fermi mechanism in shock waves in the plasma. These objects typically feature relativistic jets and shock waves can form either through interactions between the jet and its surrounding medium or through internal collisions between different jet layers \cite{1994ApJ...430L..93R}.

Simulations of jets in GRBs suggest that their Lorentz factor varies with polar angle, with a faster core jet surrounded by a slower outer layer \cite{Zhang.etal.2003ApJ...586..356Z}. This angular dependence of the Lorentz factor may be a universal feature in GRB jets \cite{Pe'er&Ryde2017IJMPD..2630018P}, potentially offering an alternative explanation for the observed power-law spectra. Instead of shock waves, this mechanism relies on the differential Lorentz factor having layers within the jet. As the GRB jets are optically thick near their bases, we demonstrate that photon scattering with electrons in the shear layers leads to a net photon energy gain. Once these photons escape, they produce a power-law spectrum. This process resembles Fermi acceleration of charged particles \cite{Blandford&Eichler1987PhR...154....1B}, where the particles gain energy through scattering with magnetic irregularities in regions of differential motion within the plasma. However, unlike the Fermi process, this mechanism involves multiple Compton scatterings of photons with electrons across different jet regions, without requiring particle acceleration.	

Thus, we propose an alternative mechanism for generating the observed power-law spectra that do not rely on charged particle acceleration or magnetic fields \cite{2023ApJ...943L...3V}. This model is applied to GRB jets to explain their high-energy power-law spectra. The model's predictions are verified through Monte Carlo simulations. It is a novel approach to estimate the jet structure from the observed GRB spectra. 

\section{Analytical model and numerical simulations}
\label{sec_model}

GRB jets are typically highly relativistic, characterized by large bulk Lorentz factors, $\Gamma \gg 1$. According to the collapsar model, GRBs are produced following the collapse of a massive star \cite{Levinson&Eichler1993ApJ...418..386L, Woosley1993AAS...182.5505W}. While the precise structure of the jet remains uncertain, various simulations of the jet eruptions from stellar interior suggest a nearly universal profile \cite{Zhang.etal.2003ApJ...586..356Z, Lundman.etal.2013MNRAS.428.2430L}. We describe the system by a spherical coordinate system ($r,\theta, \phi$) with azimuthal symmetry. The origin of this coordinate system is positioned at the centre of the star.
To analyse the jet profile, we follow these simulations and assume that the Lorentz factor of the jet varies with the polar angle ($\theta$), described by
\be 
\Gamma(\theta)= \Gamma_{\rm min}+\frac{\Gamma_{0}}{\sqrt{\left(\frac{\theta}{\theta_{\rm j}}\right)^{2p}+1}}.
\label{eq_gamma_1}
\ee 
Here $\Gamma_{\rm 0}$ and $\Gamma_{\rm min}$ represent the maximum and minimum Lorentz factor values, respectively. $\Gamma=\Gamma_0$ along the jet axis, with $\Gamma$ decreasing asymptotically according to Equation \ref{eq_gamma_1} and reaching $\Gamma=\Gamma_{\rm min}$ at large angles. The angle $\theta_j$ is fixed: for angles smaller than $\theta_j$, the jet Lorentz factor is approximately $\Gamma (\theta < \theta_j) \approx \Gamma_0$, while for angles larger than $\theta_j$, $\Gamma$ decreases as a power law with an index $p$. It represents the steepness of the shear in the jet.

We assume that the photons are emitted deep within the jet, where it is optically thick. Additionally, we consider that the Lorentz factor remains constant along the radial distance $r$. The particle density follows an inverse square law with distance (see Vyas \& Pe'er 2023 \cite{2023ApJ...943L...3V} for further details). After their emission, the photons experience multiple Compton scattering with the electrons in the shearing layers of the jet before escaping. Due to the decrease in density with radial distance, the jetted plasma becomes optically thin beyond a certain value of $r=\Rph$ (the photospheric radius). This radius is defined as the distance from the origin at which the optical depth for photon scattering from that location to infinity equals unity. Once the photons reach the photosphere, they can escape.

Below we calculate the observed spectra from the escaping photons, employing analytical methods and numerical simulations, under various assumptions regarding the uncertain jet profile.

\subsection{Analytic estimation of the observed spectra}
We assume an initial injection of $N_0$ photons at an initial energy of $\varepsilon_0$, located deep within the jet. After undergoing $k$ scattering events, we assume that $N$ photons remain inside the jet, while $(N_0-N)$ have successfully escaped. After $k$ scatterings, the average energy of a photon is $\varepsilon_k$. On average, a photon gains energy, implying that $\varepsilon_k > \varepsilon_i$. Let $\Bar{g}$ denote the average energy gain per scattering. Thus, as long as the photon remains in the scattering region, its energy after $k$ scatterings can be expressed as $\varepsilon_k = \varepsilon_i \Bar{g}^k$. Assuming an average scattering probability of $\bar{P}$, this leads to

\be 
\frac{N}{N_0} = \left[\frac{\varepsilon_k  }{\varepsilon_i}\right]^{\beta'}
\label{eq_n_n0},
\ee 
here $\beta'= \ln \bar{P}/\ln \bar{g}$. Hence, the photon index ($\beta$) is,
\be 
\beta=\beta'-1 = \frac{\ln \bar{P}}{\ln \bar{g}}-1.
\label{eq_photon_ind}
\ee
Given that the conditions within the jet change with position, the average gain and scattering probabilities are determined by calculating their expected values over the scattering volume ($V$) of the jet.
\be 
\bar{P} = \frac{1}{V}\int_{V}  P(r,\theta) dV {\rm ~ and ~} \bar{g} = \frac{1}{V}\int_{V}  g(r,\theta) dV. 
\label{eq_bar_pandg}
\ee 

The photon's escape probability at a given location is $P_{\rm e}(r,\theta) = \exp[-\tau(r,\theta)]$, where $\tau$ represents the optical depth in the direction of the photon. As the total probability must equal unity, the probability of the photon undergoing its next scattering event without escaping is given by $P(r,\theta)=1-{P_{e}(r,\theta)}$. The average energy gain at a particular location $(r,\theta)$ is estimated as [see Vyas \& Pe'er (2023) \cite{2023ApJ...943L...3V}]

\be 
g(r,\theta) \approx\frac{1}{2}\left[1+\left[1+\sum \frac{\partial \log \Gamma}{\partial \theta }\delta \theta\right]^2 \frac{1}{\left(1+a\right)^2}\right].
\label{eq_gain_at_pos}
\ee

Here, the terms $\partial \log \Gamma/\partial \theta$ represent the gain resulting from repeated scattering between the jet shear layers. $a=\lambda/r$ denotes the expansion factor that diminishes the gain due to the spherical (adiabatic) expansion of the jet, where $\lambda$ is the average mean free path at this location. For details of the estimation of $\lambda$ and $\tau$ see Vyas \& Pe'er (2023) \cite{2023ApJ...943L...3V}.
For a given set of jet parameters, $\bar{P}$ and $\bar{g}$ are calculated using equation \ref{eq_bar_pandg}, after which the photon indices of the scattered spectra at high energy are determined according to Equation \ref{eq_photon_ind}.

\subsection{Numerical simulations}
To validate the predictions of the analytic model, we conduct Monte Carlo simulations of the process outlined in the previous section. Photons are injected deep within a cold jet, with their initial directions randomly assigned in the comoving frame of the jet. For a specified density and the jet Lorentz factor profile, each photon undergoes multiple scatterings within the shear jet layer. The photons are individually tracked during each scattering event, while electrons transform their four vectors between the fluid frame, electron rest frame, and observer frame.

After each scattering, the probability for the photon to travel a distance $\delta l$ without undergoing another scattering is given by $\exp^{-\tau}$, where $\tau$ represents the optical depth, drawn from a randomly selected logarithmic distribution. If $\tau$ exceeds the optical depth required for escape to infinity, the photon leaves the system in the same direction without further scattering. Otherwise, the length $\delta l$ is determined such that the optical depth along this path equals $\tau$. As the photon propagates along the jet, the optical depth decreases in accordance with the reduction in density. This process continues until the photon escapes near the photospheric radius, where $\tau=1$.

The photons that escape are distributed across a range of energies, angular directions, and escape times, allowing the analysis of their spectral and temporal features for a given observer's location. To derive the spectrum observed by an observer, the number of photons escaping along the differential angular coordinate $\theta = \theta_o$ (the photon flux as seen by the observer) is distributed between various energy channels. Due to the symmetry along $\phi$, the spectra remain invariant with respect to the azimuthal position of the observer. The variation of this photon flux with the energy of the photons yields the observed emission spectrum for that observer. The photon indices of the observed spectrum are compared with those obtained analytically for various jet parameters. For a specific set of jet parameters, $\bar{P}$ and $\bar{g}$ are computed using equation \ref{eq_bar_pandg}, and subsequently, the photon indices of the scattered spectra at high energies are calculated according to Equation \ref{eq_photon_ind}.  Further details regarding the simulation code can be found in references \cite{Pe'er.2008ApJ...682..463P, Lundman.etal.2013MNRAS.428.2430L}
.
\section{Results}
\label{sec_results}
\begin{figure}
\includegraphics[scale=0.5] {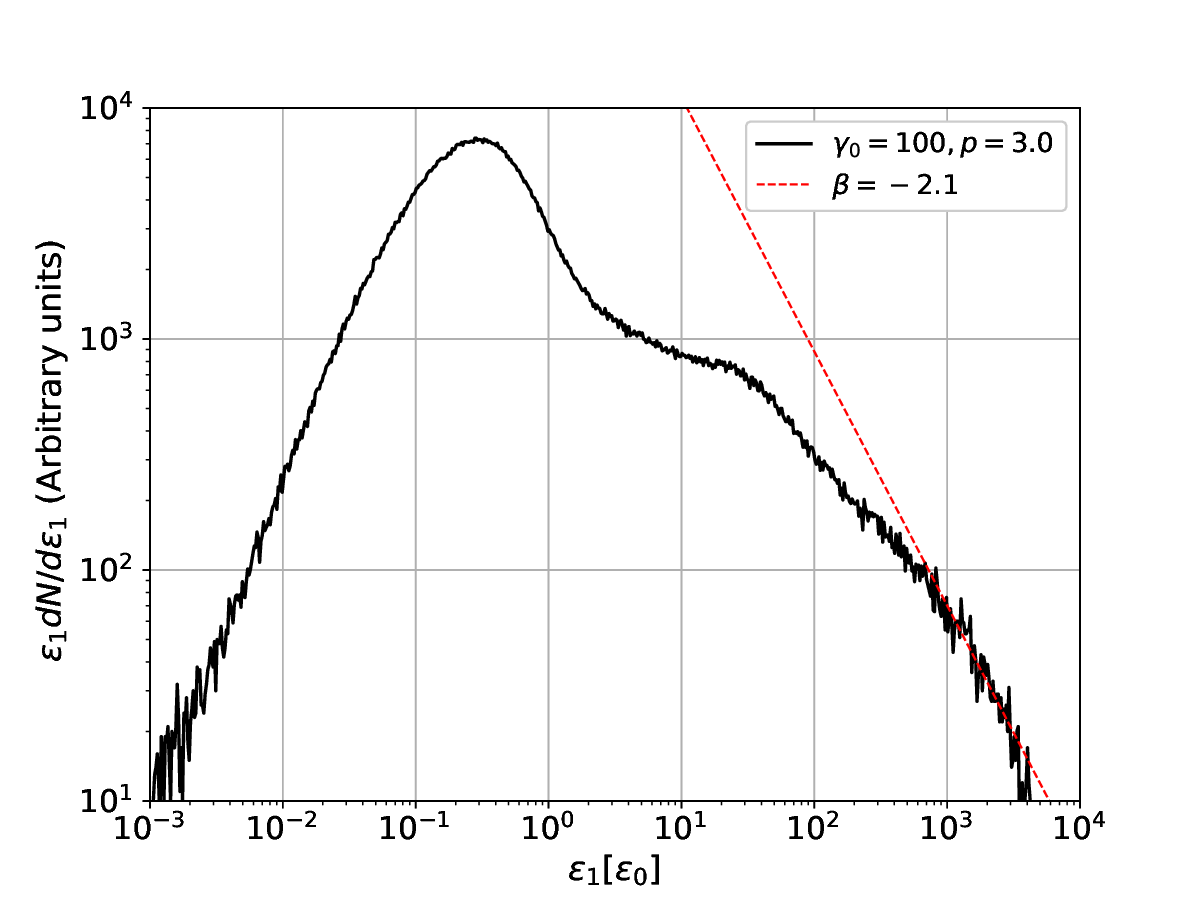}
\caption{Spectrum from numerical simulations with the observed energy $\varepsilon_1$ (black solid) and analytic spectral slope (red dashed) with parameters $L=10^{50}$ erg s$^{-1}$, $\Gamma_0 = 100, p = 3.0, \theta_j = 0.01$ rad for an on-axis observer.} 
\label{label_spectrum_p3.0}
\end{figure}

We consider a shearing jet characterized by the following parameters: luminosity $L=10^{50}$ erg s$^{-1}$, $\Gamma_0=100$, $\tj=0.01$ rad, $\gmin=1.1$, and $p=3$. Photons are injected deep within the jet, with energy $\varepsilon_0 = 10^{-7}$ in units of the electron's rest energy. The observed spectrum across the observed energy range $\varepsilon_1$ is derived from numerical simulations, while the high-energy slope is calculated analytically. The binned spectrum from our simulations is illustrated in Figure \ref{label_spectrum_p3.0} for an observer located along the jet axis (black solid curve). This spectrum exhibits a power-law shape at energies $\varepsilon>5 \times 10^2 \varepsilon_0$. The analytically computed photon index (as per Equation \ref{eq_photon_ind}) for the specified parameters is $\beta = -2.1$. This is over-plotted with the red dashed curve on the simulated spectra. The analytic prediction of the photon index aligns with the photon index obtained through numerical simulations validating the theory.

\section{Conclusions}
\label{sec_conclusions}
Here we have examined a shearing jet with a variable Lorentz factor dependent on its polar angle. We demonstrated that a photon population injected deep within the flow and scattering in an optically thick shearing jet experiences an average energy gain, resulting in a power-law spectrum at high energies. The theoretical model developed is in strong agreement with the simulation results. The chosen parameters for a typical GRB jet produces a power-law index in the range $\beta \rightarrow -\infty$ to $-1.5$. This index aligns well with the observed range of high-energy spectral slopes ($\beta$) during the prompt phase of GRBs \cite{Preece.etal.2000ApJS..126...19P, Kaneko.etal.2006ApJS..166..298K, pe'er2015AdAst2015E..22P}. Therefore, we conclude that the alternative mechanism proposed here can sufficiently reproduce the observed GRB spectra at high energies.

The model also has applications beyond GRB spectra, including in active galactic nuclei (AGNs) \cite{2024ApJ...972...40V}. Further applications of this model will be explored and published in future.

\section*{Acknowledgements}
We are thankful to the European Union
(EU) via ERC consolidator grant 773062 (O.M.J.). 
\bibliography{sample}{}

\providecommand{\noopsort}[1]{}\providecommand{\singleletter}[1]{#1}%
\begin{thebibliography}{10}

\bibitem{Band.etal.1993}
D.~{Band}, J.~{Matteson}, L.~{Ford}, B.~{Schaefer}, D.~{Palmer},
  B.~{Teegarden}, T.~{Cline}, M.~{Briggs}, W.~{Paciesas}, G.~{Pendleton},
  G.~{Fishman}, C.~{Kouveliotou}, C.~{Meegan}, R.~{Wilson} and P.~{Lestrade},
  {\em ApJ} {\bf 413} (August 1993)   281.

\bibitem{Kaneko.etal.2006ApJS..166..298K}
Y.~{Kaneko}, R.~D. {Preece}, M.~S. {Briggs}, W.~S. {Paciesas}, C.~A. {Meegan}
  and D.~L. {Band}, {\em ApJs} {\bf 166} (September 2006) 298,
  \href{http://arxiv.org/abs/astro-ph/0601188}{{\ttfamily
  arXiv:astro-ph/0601188 [astro-ph]}}.

\bibitem{Bosmjak.etal.2014A&A...561A..25B}
{\v{Z}}.~{Bo{\v{s}}njak}, D.~{G{\"o}tz}, L.~{Bouchet}, S.~{Schanne} and
  B.~{Cordier}, {\em A\&A} {\bf 561} (January 2014)   A25,
  \href{http://arxiv.org/abs/1309.3174}{{\ttfamily arXiv:1309.3174
  [astro-ph.HE]}}.

\bibitem{pe'er2015AdAst2015E..22P}
A.~{Pe'er}, {\em Advances in Astronomy} {\bf 2015} (January 2015)   907321,
  \href{http://arxiv.org/abs/1504.02626}{{\ttfamily arXiv:1504.02626
  [astro-ph.HE]}}.

\bibitem{Barraud.etal.2003A&A...400.1021B}
C.~{Barraud}, J.~F. {Olive}, J.~P. {Lestrade}, J.~L. {Atteia}, K.~{Hurley},
  G.~{Ricker}, D.~Q. {Lamb}, N.~{Kawai}, M.~{Boer}, J.~P. {Dezalay},
  G.~{Pizzichini}, R.~{Vanderspek}, G.~{Crew}, J.~{Doty}, G.~{Monnelly},
  J.~{Villasenor}, N.~{Butler}, A.~{Levine}, A.~{Yoshida}, Y.~{Shirasaki},
  T.~{Sakamoto}, T.~{Tamagawa}, K.~{Torii}, M.~{Matsuoka}, E.~E. {Fenimore},
  M.~{Galassi}, T.~{Tavenner}, T.~Q. {Donaghy}, C.~{Graziani} and J.~G.
  {Jernigan}, {\em A\&A} {\bf 400} (March 2003) 1021,
  \href{http://arxiv.org/abs/astro-ph/0206380}{{\ttfamily
  arXiv:astro-ph/0206380 [astro-ph]}}.

\bibitem{Nandra&Pounds1994}
K.~{Nandra} and K.~A. {Pounds}, {\em MNRAS} {\bf 268} (May 1994) 405.

\bibitem{Page.etal.2005}
K.~L. {Page}, J.~{Reeves}, P.~T. {O'Brien} and M.~J.~L. {Turner}, {\em MNRAS}
  {\bf 364} (November 2005) 195,
  \href{http://arxiv.org/abs/astro-ph/0508524}{{\ttfamily
  arXiv:astro-ph/0508524 [astro-ph]}}.

\bibitem{1970RvMP...42..237B}
G.~R. {Blumenthal} and R.~J. {Gould}, {\em Reviews of Modern Physics} {\bf 42}
  (January 1970) 237.

\bibitem{1986rpa..book.....R}
G.~B. {Rybicki} and A.~P. {Lightman}, {\em {Radiative Processes in
  Astrophysics}} 1986.

\bibitem{1993ApJ...409L..33Z}
A.~A. {Zdziarski} and J.~H. {Krolik}, {\em ApJl} {\bf 409} (June 1993)   L33.

\bibitem{vyas.etal.2021.predictingAPJL}
M.~K. {Vyas}, A.~{Pe'er} and D.~{Eichler}, {\em ApJl} {\bf 918} (September
  2021)   L12, \href{http://arxiv.org/abs/2103.06201}{{\ttfamily
  arXiv:2103.06201 [physics.space-ph]}}.

\bibitem{1994ApJ...430L..93R}
M.~J. {Rees} and P.~{Meszaros}, {\em ApJl} {\bf 430} (August 1994)   L93,
  \href{http://arxiv.org/abs/astro-ph/9404038}{{\ttfamily
  arXiv:astro-ph/9404038 [astro-ph]}}.

\bibitem{Zhang.etal.2003ApJ...586..356Z}
W.~{Zhang}, S.~E. {Woosley} and A.~I. {MacFadyen}, {\em ApJ} {\bf 586} (March
  2003) 356, \href{http://arxiv.org/abs/astro-ph/0207436}{{\ttfamily
  arXiv:astro-ph/0207436 [astro-ph]}}.

\bibitem{Pe'er&Ryde2017IJMPD..2630018P}
A.~{Pe'er} and F.~{Ryde}, {\em International Journal of Modern Physics D} {\bf
  26} (January 2017) 1730018, \href{http://arxiv.org/abs/1603.05058}{{\ttfamily
  arXiv:1603.05058 [astro-ph.HE]}}.

\bibitem{Blandford&Eichler1987PhR...154....1B}
R.~{Blandford} and D.~{Eichler}, {\em Physics Reports} {\bf 154} (October 1987)
  1.

\bibitem{2023ApJ...943L...3V}
M.~K. {Vyas} and A.~{Pe'er}, {\em ApJl} {\bf 943} (January 2023)  ~L3,
  \href{http://arxiv.org/abs/2207.11481}{{\ttfamily arXiv:2207.11481
  [astro-ph.HE]}}.

\bibitem{Levinson&Eichler1993ApJ...418..386L}
A.~{Levinson} and D.~{Eichler}, {\em ApJ} {\bf 418} (November 1993)   386.

\bibitem{Woosley1993AAS...182.5505W}
S.~E. {Woosley}, { {Gamma-Ray Bursts from Stellar Collapse to a Black Hole?}},
  in {\em American Astronomical Society Meeting Abstracts \#182\/}, , American
  Astronomical Society Meeting Abstracts Vol.~182 (May 1993), p. 55.05.

\bibitem{Lundman.etal.2013MNRAS.428.2430L}
C.~{Lundman}, A.~{Pe'er} and F.~{Ryde}, {\em MNRAS} {\bf 428} (January 2013)
  2430, \href{http://arxiv.org/abs/1208.2965}{{\ttfamily arXiv:1208.2965
  [astro-ph.HE]}}.

\bibitem{Pe'er.2008ApJ...682..463P}
A.~{Pe'er}, {\em ApJ} {\bf 682} (July 2008) 463,
  \href{http://arxiv.org/abs/0802.0725}{{\ttfamily arXiv:0802.0725
  [astro-ph]}}.

\bibitem{Preece.etal.2000ApJS..126...19P}
R.~D. {Preece}, M.~S. {Briggs}, R.~S. {Mallozzi}, G.~N. {Pendleton}, W.~S.
  {Paciesas} and D.~L. {Band}, {\em ApJs} {\bf 126} (January 2000) 19,
  \href{http://arxiv.org/abs/astro-ph/9908119}{{\ttfamily
  arXiv:astro-ph/9908119 [astro-ph]}}.

\bibitem{2024ApJ...972...40V}
M.~K. {Vyas} and A.~{Pe'er}, {\em ApJ} {\bf 972} (September 2024)  ~40,
  \href{http://arxiv.org/abs/2308.06728}{{\ttfamily arXiv:2308.06728
  [astro-ph.HE]}}.

\end{thebibliography}
\bibliographystyle{ws-ijmpd}
%
%
%
%
%
%
%

\end{document}